\documentclass[9pt,twocolumn,twoside]{osajnl}
\usepackage{soul} 

\journal{ol} 
\setboolean{shortarticle}{true}
\title{100,000 frames-per-second compressive imaging with a conventional rolling-shutter camera by random point-spread-function engineering}

\author[ ]{Gil Weinberg}
\author[*]{Ori Katz}

\affil[ ]{Department of Applied Physics, Hebrew University of Jerusalem, Jerusalem 9190401, Israel}
\affil[*]{orik@mail.huji.ac.il}

\begin{abstract}
We demonstrate an approach that allows taking videos at very high-speeds of over 100,000 frames per second (fps) by exploiting the fast sampling rate of the standard rolling-shutter readout mechanism, common to most conventional sensors, and a compressive-sampling acquisition scheme.  Our approach is directly applied to a conventional imaging system by the simple addition of a diffuser to the pupil plane, randomly encoding the entire field-of-view to each camera row, while maintaining diffraction-limited resolution. A short video is reconstructed from a single camera frame via a compressed-sensing reconstruction algorithm, exploiting inherent sparsity of the imaged scene. 
\end{abstract}

\setboolean{displaycopyright}{false} 

\begin{document}

\maketitle
    High-speed imaging is important for observations of fast occurring phenomena, such as high-speed tracking of single-particles \cite{shen2017single}, the study of explosive materials, in the automotive industry, and more \cite{mikami2016ultrafast}. While high-resolution cameras are available today in affordable devices such as in smartphones, very high-speed ($>$10,000 frames-per-second (fps)) cameras are still uncommon and expensive, with price tags at the thousands of dollars. The high price-tag is a result of the hardware requirements, which include high-speed readout, a high transfer bit-rate from the sensor to the memory, and a high-speed memory allocated close to the sensor.Here, we suggest a simple solution for high-speed (>100,000fps) imaging of sparse-scenes that goes beyond the hardware transfer bit-rate limit, by exploiting the fast rolling-shutter readout common to most conventional sensors, and a compressive-sampling (CS) acquisition scheme.
    
    Our technique can be applied to any conventional imaging system by the simple addition of a diffuser at the imaging-system pupil-plane. The diffuser goal is to generate a random speckle point-spread-function (PSF), which encodes the entire scene to each camera row. The rolling shutter samples the scene at 100,000 rows-per-second, allowing high-speed video reconstruction from a single camera frame.
    
    In recent years, various techniques that allow video reconstruction from a single frame based on compressed-sensing have been developed \cite{liang2018single}. Among them are coded exposure photography \cite{raskar2006coded}, dynamic coded-aperture photography \cite{llull2013coded}, dynamic point-spread-function (PSF) modulation \cite{shi2019speckle} and many more \cite{gao2014single,koller2015high, gu2010coded,hitomi2011video, sun2017compressive,yuan2016compressive}. However, these approaches rely on a fast dynamic modulation, and require relatively-complex modifications or additions to the imaging system.

    In contrast to these approaches, our approach requires only a simple, straightforward, addition of a static diffuser to the imaging system pupil-plane. Recently, a similar technique of utilizing the rolling-shutter effect for video reconstruction from a single frame was presented by Antipa et al. \cite{antipa2019video} for lensless imaging systems. The implementation of this approach, however, required access to the bare sensor, and thus, is not directly adaptable to all conventional imaging systems. Here, we demonstrate that the use of the rolling-shutter for high-speed photography can be exploited in any lens-based imaging system by accessing only the pupil plane. In addition, by relying on a speckle-based encoding, rather than a caustics-one used by Antipa et al., our implementation preserves a diffraction-limited spatial resolution, a potential advantage for microscopic imaging.

    The enabling principle in all approaches of video reconstruction from a single frame is that the dynamic scene to be captured is compressible in some known domain and that each pixel in the acquired frame encodes information from a large number of video pixels, in both \textit{space} and \textit{time}. Following the principles of CS \cite{donoho2006compressed}, a dynamic video can be reconstructed from such a single frame under the conditions of an appropriate encoding and a minimal number of acquired pixels. Importantly, a random encoding, i.e. an encoding where each camera pixel is a random linear superposition of video pixels in space and time, satisfies those conditions.
    
\begin{figure}[hbt!]
    \centering
    \includegraphics[width=\linewidth,]
    {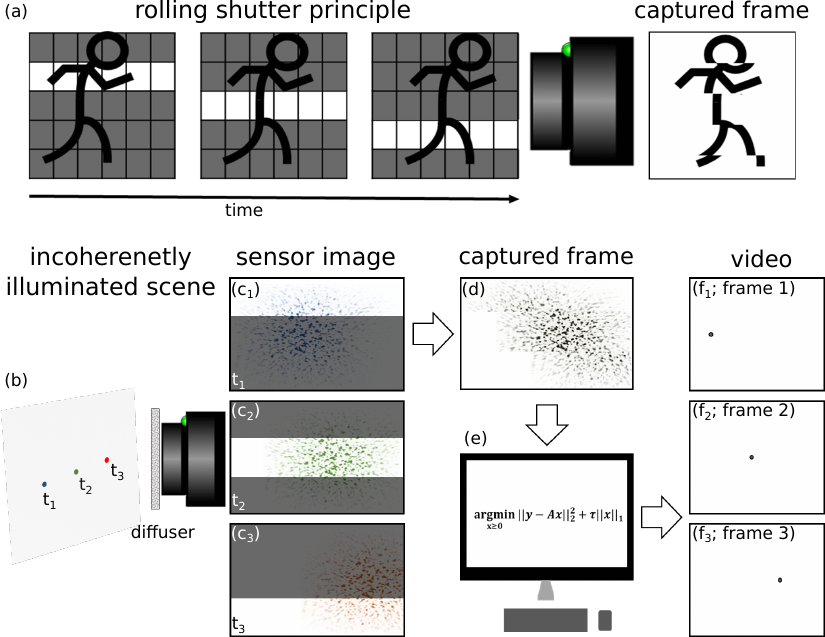}
    \caption{ 
       Principle. (a) In a rolling-shutter readout, the sensor rows are sampled consecutively at different times, yielding a sampling-rate that is orders of magnitude faster than the full frame-rate. (b) The fast sampling-rate is exploited for video capture by spreading the light from each point in the imaged scene (left) over all camera rows, by an optical diffuser placed at the camera pupil plane, generating a random speckle PSF. (b-f) An illustrated example of imaging a rapidly moving point source: (c) at each time $t_i$ the moving source projects the speckle PSF at a different position on the sensor, and a different row (in white) is sampled. Each row in the full captured camera frame (d) reflects the scene at a different time. (e) a compressed-sensing algorithm reconstructs the video (f) from the single captured camera frame.
        }
    \label{method}
\end{figure}

    In a conventional camera with a rolling shutter readout, the camera image is sampled row after row in a serial, consecutive manner \cite{gu2010coded}. Therefore, each row of pixels in the acquired frame encodes the scene information in a specific, short, time (Fig.\ref{method}a). This effect is usually a hurdle for high-speed photography \cite{liang2008analysis}, causing artifacts for fast-moving objects, or fast-changing scenes, as is depicted in Fig. \ref{method}a. However, one can exploit the inherently fast sampling-rate of the rolling shutter for video reconstruction by randomly-encoding the entire scene information (in all spatial and temporal coordinates) to every camera row (scanline).
    
    A schematic realization of this principle using an optical diffuser is presented in Fig.\ref{method}b-g. A light scattering diffuser at the pupil plane of the imaging system (Fig.\ref{method}c) optically randomly encodes the entire scene to each row in the camera sensor by scattering (Fig.\ref{method}d). A camera with a rolling-shutter readout captures the single frame (Fig.\ref{method}e). Each row in the captured image encodes a single time frame of the video. The single captured image is fed into a CS reconstruction algorithm that decodes the video (Fig.\ref{method}f-g).
    Importantly, in common CMOS cameras, the readout speed of each row is usually several orders of magnitude faster than the full image acquisition rate (the number of rows in the image) \cite{gu2010coded}.

    The acquisition and reconstruction of our approach can be mathematically described as follows: for imaging of a 2D spatially-incoherently illuminated scene, described by $O(x,y,t)$, the image intensity at the sensor plane at any time, $t$, is given by: 
    \begin{equation}\label{Incoherent Imaging}
        I(x,y,t) = O(x/M,y/M,t) \underset{x,y} \circledast  PSF(x,y)
    \end{equation}
    Where PSF(x,y) is the imaging PSF, $M$ is the magnification of the imaging system, and $\underset{x,y} \circledast$ denotes 2D spatial-convolution over (x,y).
    
    The single image acquired by a camera with a rolling-shutter readout is a sampled version of $I(x,y,t)$, where the row at the coordinate $y$ is sampled at a time, $t=y/V_s$, where $V_s$ is the rolling shutter speed. Assuming an exposure time given by $T_{exp}$ the captured image is given by:
    \begin{equation}\label{Rolling Shutter Imaging}
        I_{cam}(x,y) = \int^{\frac{y}{V_s}}_{\frac{y}{V_s}-T_{exp}} \ I(x,y,t)dt 
    \end{equation}
   When the exposure time is shorter than the time it takes the rolling shutter to pass through a single row: $T_{exp}\leq D_{row}/V_s$, where $D_{row}$ is the height of each pixel, \eqref{Rolling Shutter Imaging} can be approximated by:
       \begin{equation}\label{Rolling Shutter Imaging approx}
            I_{cam}(x,y) \approx O(x,y,\frac{y}{V_s}) \underset{x,y} \circledast PSF(x,y)
       \end{equation}
   This simple approximation holds if the scene dynamics are slower than the exposure time, and allows temporal discretization with a temporal resolution of $T_{exp}$. Finer temporal discretization is also possible, but for the sake of simplicity we will not consider it here.
    
    The linear forward model described in \eqref{Rolling Shutter Imaging approx} can be written as a matrix-vector multiplication of a convolution-like matrix, $\textbf{A}$ that describes the PSF and the rolling-shutter readout, with a vector $\textbf{b}$ that describes the dynamic scene $O(x,y,t)$ \cite{antipa2018diffusercam,pascucci2019compressive}.
    
    \begin{equation}\label{Forward Model}
        \textbf{b} = \textbf{A}\textbf{v} 
    \end{equation}
    Where $\textbf{b}$ is a vector of dimensions $[m=n_x\cdot n_y,1]$, representing the captured $n_x$ by $n_y$ pixels-image. $\textbf{v}$ is a vector with dimensions $[n = n_x \cdot n_y \cdot n_t,1]$ that represents the 3D dynamic scene having $n_t$ temporal bins, and $\textbf{A}$ is the forward-model (\eqref{Rolling Shutter Imaging approx}) matrix of dimensions $[m,n]$. The $i$-th row in the matrix $\textbf{A}$ describes the random-encoding of the spatio-temporal scene to the $i$-th sensor-pixel. The $j$-th column in $\textbf{A}$ describes the spreading of each spatio-temporal pixel intensity in the scene $I(x_j,y_j,t_j)$ to all camera pixels. The columns in $\textbf{A}$ are thus shifted, sampled representations of the PSF.
    
    In the framework of CS, the scene video, $\textbf{v}$, can be reconstructed from the acquired frame, $\textbf{b}$, by e.g. finding the solution to the convex-minimization problem \cite{donoho2006compressed}:
    
    \begin{equation}\label{inverse problem}
        \tilde v = \underset{v \geq 0}{\operatorname{argmin}} ||b-Av|^2_2 + \tau ||\Psi v||_1
    \end{equation}
    where $\textbf{$\Psi$}$ is a linear transformation matrix mapping $\textbf{v}$ to a domain where it has a sparse representation. For example $\textbf{$\Psi$}$ can be a spatial discrete cosine-transform \cite{katz2009compressive}, a spatial and/or temporal derivative aimed at minimizing the spatio-temporal total variation \cite{chan2011augmented}.  $\tau$ is a regularization parameter chosen according to the scene sparsity and measurement signal-to-noise (SNR). A large $\tau$ should be chosen for a sparse scene and/or a low SNR.

    According to CS theory, for a scene represented by a $k$-sparse representation, a high-fidelity reconstruction is possible if the number of measurements, $m$, satisfies: $m \geq \mathcal{O}(k \cdot log(n))$ \cite{romberg2008imaging}. In our case, $m$ is the number of camera pixels, and $n$ is the number of spatio-temporal pixels in the video.
    
     The imaging procedure requires performing three steps: PSF calibration, acquisition, and reconstruction. For calibration, a point-object is used to record the system PSF. For an ideal thin diffuser placed at the pupil plane, the system is expected to be isoplanatic and a single PSF recording is sufficient. For realistic diffusers, the isoplanatic angle is the angular "memory effect" \cite{freund1988memory} of the diffuser, which for ground-glass or holographic diffusers is a couple of degrees \cite{katz2012looking,bertolotti2012non}. Thus, the PSF calibration can be done by recording the PSF in a few isoplanatic patches. The acquisition (encoding) step is a simple recording of a single camera frame. Finally, the reconstruction (decoding) step is performed by running a conventional CS reconstruction algorithm to solve the inverse problem \cite{figueiredo2007gradient}. 

\begin{figure}[!hbt]
    \centering
    \includegraphics[width=\linewidth]
    {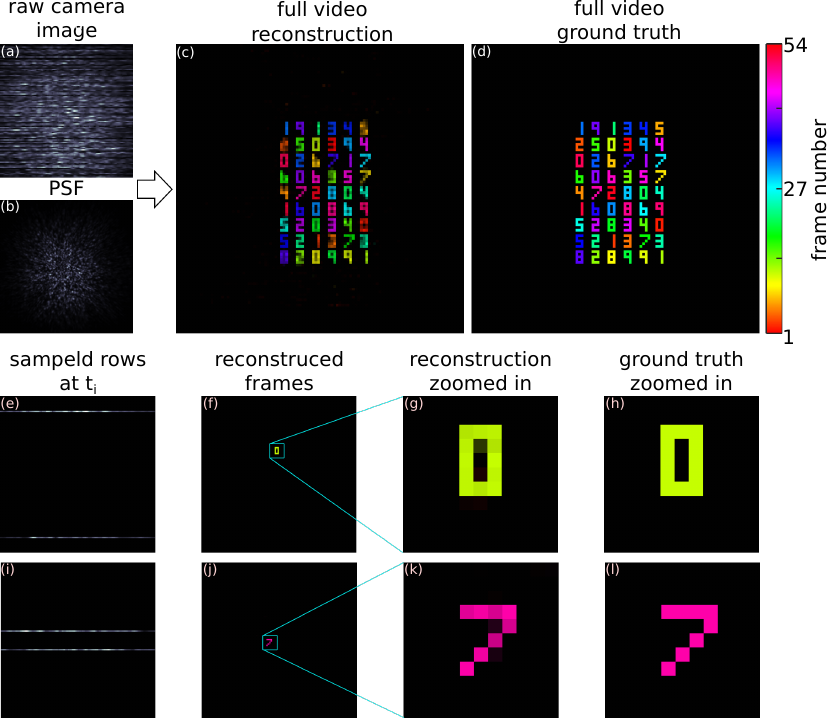}
    \caption{
            Numerical simulation: Acquisition of a dynamic scene by a rolling shutter sensor with a pixel resolution of 108x108 (a) is simulated using an experimentally measured speckle PSF (b), and an SNR=256. The dynamic scene is composed of 54 time bins with temporally and spatially changing digits. The reconstructed short video (c) is in excellent agreement with the ground truth (d). The different colors represent frames at different times. (e-h) reconstruction of frame 11 from the video: (e) captured rows of the $11^{th}$ time bin, (f) reconstructed frame, (g) zoom-in on (f), (h) same as (g) ground-truth for (g). (i-l) same as (e-h) for the $48^{th}$ frame.
            }   
            
    \label{simulation_results}
\end{figure}

    As a first step to confirm the proposed approach, we performed a numerical simulation using an experimentally measured PSF, generated by 1° diffuser (Newport). The results of this simulation are presented in Fig. \ref{simulation_results}. The simulated scene is composed of 54 digits changing in space and time over 54 time bins and 108x108 spatial pixels. The captured image contains 108x108 pixels. Thus, m=11,644, n=629,856. Gaussian random noise was added to the raw image pixels, to simulate a measurement SNR of 256. To appropriately simulate the dual rolling shutter of our camera (Andor Zyla 4.2), two camera rows are simultaneously sampling the scene at each specific time.Under these conditions, a high-fidelity reconstruction of the 54 video frames at a resolution of 108x108 pixels was obtained (Fig. \ref{simulation_results}, yielding a $\times 54$ increase in acquisition speed, compared to the raw camera frame-rate. 

\begin{figure}[hbt!]
    \centering
    \includegraphics[width=\linewidth]
    {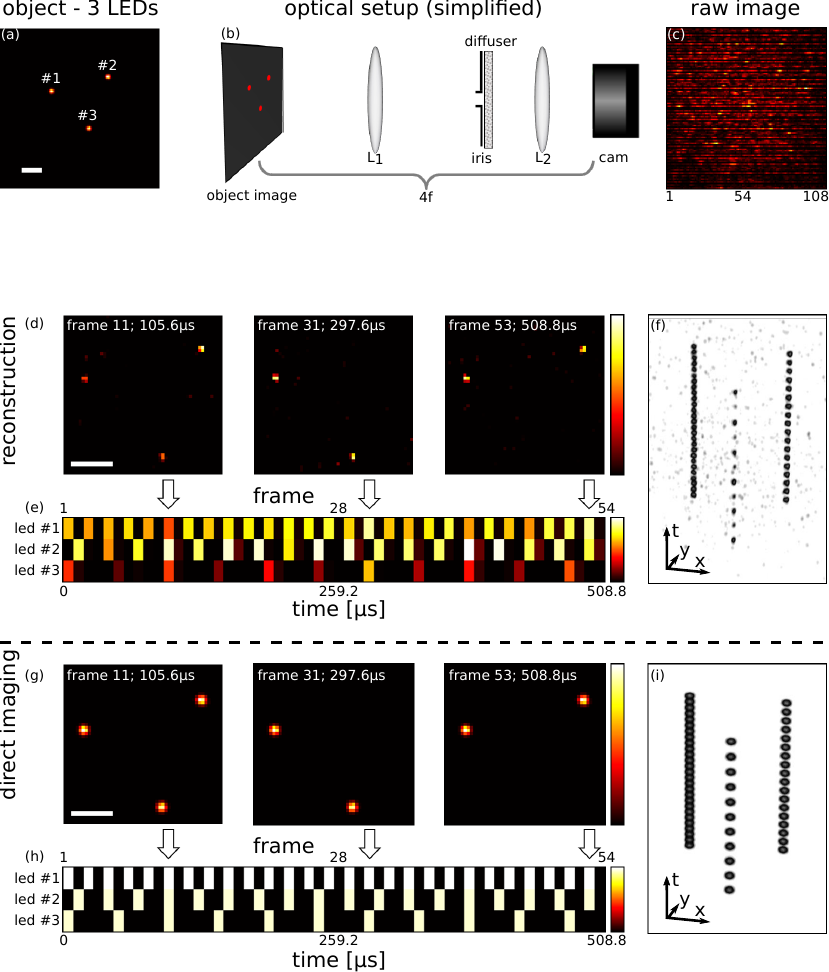}

        \caption{
            Experimental proof of principle. (a) the imaged scene, composed of three fast modulated LEDs. (b) The optical setup is an 8-f imaging system with a 1° optical diffuser at the Fourier plane (4-f shown for simplicity). (c) the raw image is captured at a pixel resolution of 108x108, and a $9.6\mu s$ exposure time. (d-f) a 54-frame video is reconstructed from the single frame at a frame rate of 104,166 fps, at the same 108x108 pixel resolution. (d) three reconstructed frames. (e) time traces for three pixels at the LEDs locations. (f) spatio-temporal presentation of the reconstructed video. (h-j) same as (d-f) for direct, diffraction-limited imaging (d), and (i) temporal traces recorded with a fast photodiode. Scale bars: 2 mm. 
        }
    \label{setup_results}
\end{figure}

    For a proof-of-principle experimental demonstration the setup depicted in Fig.\ref{setup_results}b was constructed. The setup is a simple imaging setup consisting of two 4f imaging telescopes with a $\times22.5$ de-magnification. An optical diffuser (10DKIT-C1 1° Newport) without a dominant ballistic component (zero-order diffraction peak) is placed at the Fourier plane (Fig.  \ref{setup_results}b). 
    The criterion for choosing the scattering angle of the optical diffuser is that the PSF of a point object located at the bottom of the field-of-view will reach the top row of the camera sensor. A larger PSF would be wasteful for the photon flux reaching the sensor. A smaller PSF would not encode the information from the lower and upper edges of the field-of-view to all camera rows.

    As a dynamic, rapidly changing scene, three LEDs at a wavelength of $625nm$  (M625F2, fiber coupled to 200$\mu$m fibers) were modulated at different frequencies up to 52KHz (20,833Hz, 34,722Hz, and 52,083Hz)  by three independent function generators.  
    An sCMOS camera (Andor Zyla 4.2) set to fast line readout mode and an exposure time of 9.6$\mu$s was used to capture the scene. 
    A sample video with 54 frames reconstructed at a frame rate of 104,166 fps at a pixel resolution of 108x108 pixels is presented in Fig. \ref{setup_results}. The reconstructed frame-rate is $\times 60$ higher than the highest native frame-rate of the camera at such a small ROI, and $\times 1000$ higher than the camera frame-rate at full resolution. 
    
    We presented a method for high-speed imaging that relies on the rolling shutter effect. While placing an optical diffuser at the pupil plane results in a random speckle PSF, it does not affect the resolution limit of the imaging system, since the speckle grain-size  (Fig. \ref{simulation_results}b) is diffraction-limited \cite{goodman2007speckle}.  While in our demonstration we used a simple diffuser that produces a speckle pattern PSF obeying Rayleigh-statistics \cite{goodman2007speckle}, an engineered phase-mask can be used to customize more efficient encoding or different intensity statistics \cite{bender2018customizing}. 

    As speckles can be very sensitive to the optical wavelength \cite{redding2013compact},  broadband scenes may result in low contrast raw images. This can be alleviated by engineered phase masks. However, the spectral sensitivity of speckles can be an interesting advantage for hyperspectral (x,y,t,$\lambda$)  imaging \cite{golub2016compressed, sahoo2017single}. Moreover, the suggested approach could also be extended for recovering depth information (x,y,t,z) by exploiting the natural orthogonality of speckles at different axial planes, as was recently demonstrated \cite{pascucci2019compressive}. However, both of those extensions would come at a cost of more demanding sparsity constraints since a higher dimensional vector should be reconstructed from the same number of measurements (camera pixels). To maximize reconstruction fidelity the PSF can be specifically engineered, as was recently demonstrated for the goal of 3D reconstruction using deep-learning \cite{nehme2019dense}.
    
    As in similar approaches, an inherent drawback of the presented approach is in the fact that the intensity from each spatial position in the scene is spread over a large number of camera pixels, reducing the raw signal to noise. The approach is thus not ideal for high pixel-count imaging of low photon-flux scenes.  
    For our demonstration we have used a CS reconstruction algorithm. However, a deep-learning based reconstruction framework can also be used \cite{Qiao2020Deep} for improving the reconstruction run-time and potentially the reconstruction fidelity.
    Finally, the approach can also be extended to spatially-coherent scenes, e.g. for high-speed holographic video \cite{wang2017compressive}, or in optical coherence tomography (OCT) \cite{huang1991optical}, by considering holographic detection of the fields rather than intensity only detection. 
 
    \medskip
    \noindent\textbf{Funding.} This work is funding by the organizations: European Research Council (ERC) Horizon 2020 research and innovation program (grant no. 677909), Azrieli foundation, Israel Science Foundation (1361/18), Israeli Ministry of Science and Technology.

    \bibliography{references}

    \bibliographyfullrefs{references}

\end{document}